# Who asks questions at astronomy meetings?

*Sarah J. Schmidt and James R. A. Davenport*

Over the last decade, significant attention has been drawn to the gender ratio of speakers at conferences, and many meetings are evolving to better reflect the gender balance of the field as a whole. We find that women are significantly under-represented, however, among the astronomers asking questions after talks.

Conferences act as a microcosm for our field, with the same biases and barriers to inclusion that affect the careers of astronomers played out in person instead of in abstract. The networking that occurs at conferences is essential to career development, and the in-person interactions between researchers have the potential to shape entire subfields. Recently there have been strong gains in the fraction of women speaking at conferences, as seen in the data collected from 2008 to 2013 by the Committee on the Status of Women in Astronomy (https://cswa.aas.org/percent.html). However, less is known about the nature and equity of the actual interactions scientists have at these conferences, such as in the question-and-answer sessions following each conference talk. This prompted us to conduct a volunteer-driven study of gender ratios in conference talks. Now in its fourth year, this ongoing survey has revealed many interesting biases in conference talks and questions.

**Gathering Data on Gendered Questions**

During the 223rd meeting of the American Astronomical Society (AAS; January 2014), we first began to gather data on the gender of people asking questions after talks. The project was announced via Twitter, with a short explanation of the basic goals as well as a link to a simple online form optimized for both computer and phone browsers. This form asked voluntary and anonymous participants to identify the session, the perceived gender of the speaker, and the perceived gender of each person asking questions. The results from AAS 223, examined during the AAS "Hack Day" event, confirmed the hypothesis that men are over-represented in conference questions compared to speakers or overall conference attendees (Davenport et al. 2014).

To date, a version of the gender question survey has been deployed at more than 12 conferences, including five AAS meetings. So far data from the initial AAS 223 (Davenport et al. 2014), the 2014 UK National Astronomy Meeting (Pritchard et al.

2014) and Cool Stars 18 and 19 (Schmidt et al. 2017) have been analyzed in detail. The most complete set of data so far is a compilation of responses from the four successive AAS winter meetings in 2014–2017 (223, 225, 227, and 229), each with responses from at least 150 talks. In total the dataset samples nearly 1000 talks, and over 2500 questions.

The overall results, shown in Figure 1, are consistent between all four AAS meetings. The gender ratio of speakers (36% women and 64% men) typically mirrors the ratio of attendees (35% women and 65% men at AAS 223; Davenport et al. 2014), indicating the speakers are drawn fairly from the astronomers in attendance. The attendees are, on average, more female than the overall AAS membership (73% male and 25% female), and instead may be drawn from a more junior portion of the field (AAS members born after 1980 are 60% male and 40% female; Anderson & Ivie 2014). The gender ratio of people asking questions did not mirror the ratio of attendees, with men asking 75% of questions while women asked 25%. If we assume the data gathered is a representative sampling of the entire conference, each male AAS attendee asks an average of 0.93 questions per meeting, while each female attendee asks 0.57 (based on data from the AAS 223 attendees).

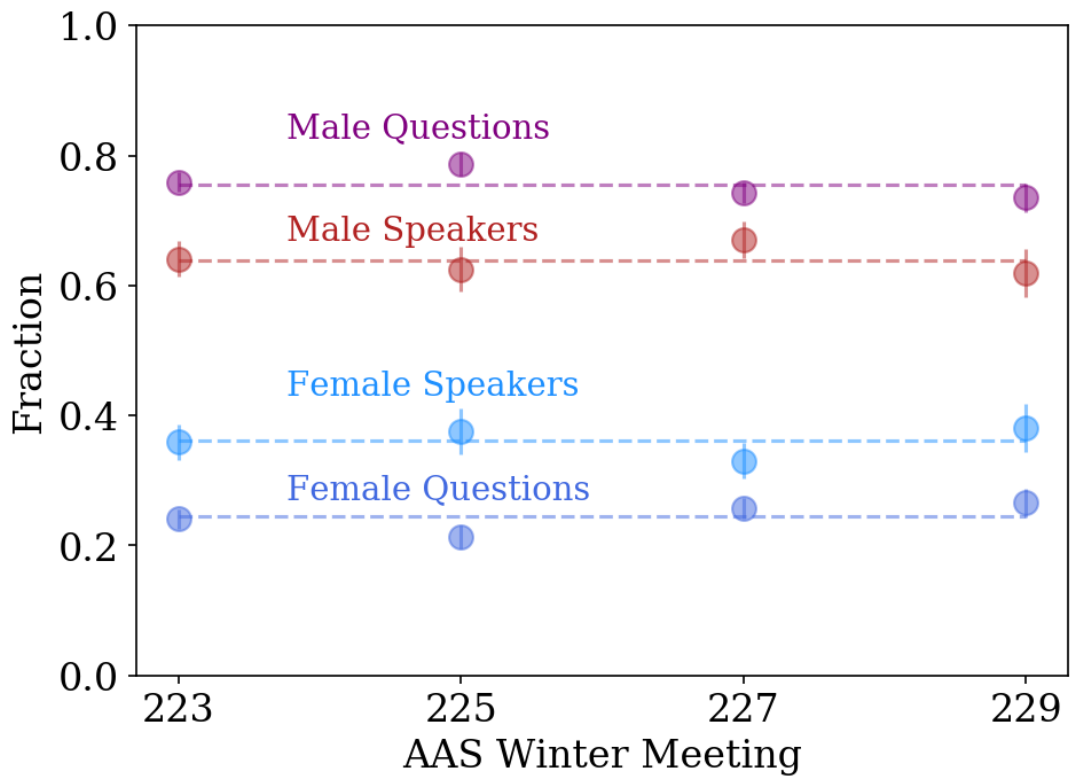

Figure 1 | The fraction of male and female questioners and speakers at four successive winter AAS meetings. In addition to the data from individual meetings (with Poisson uncertainties shown), the mean for all four meetings of male questions (76%), male speakers (64%), female speakers (36%), and female questions (24%) are indicated (dashed lines), with standard errors on the mean of ~0.01 for each. The results from individual meetings show no significant change over four years.

**Sharing the Air**

One intriguing result (initially found by Pritchard et al. 2014) is a correlation between the total number of questions asked after a given talk and the resulting fraction of questions asked by women. As shown in Figure 2, we detect this correlation in the AAS meeting data as well, and are able to see in detail the impact of Q/A session length. In talks with only one or two questions the fraction of questions from women was only 24%. As the number of questions increases, the fraction of questions asked by women steadily increases until for talks with six or more questions it mirrors the attendance gender ratio. While the data indicate a clear relationship, the many contributing factors are impossible to determine from our data. For example, it is possible that women on average think about their questions for longer, while men raise their hands quickly. It is also possible that session chairs hold subtle biases towards male scientists and call on them first or faster; these biases have been documented in other areas (see, e.g., the resume comparison work by Moss-Racusin et al. 2012).

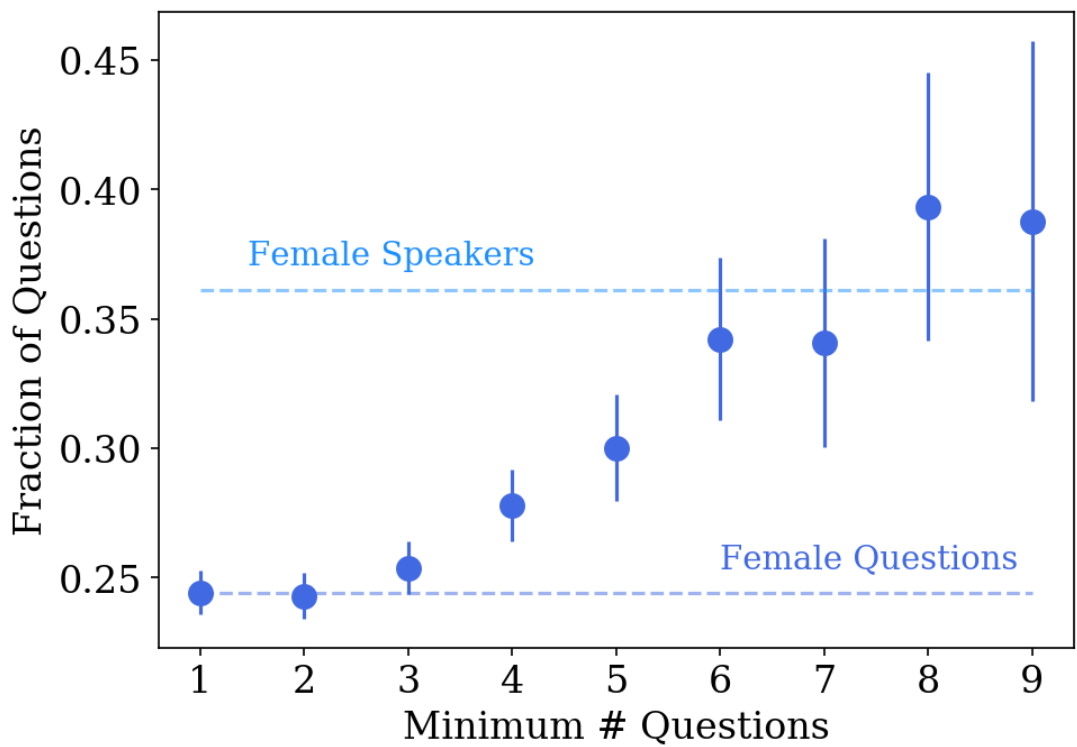

Figure 2 | The fraction of questions asked by women as a function of the total number of questions taken during the session. The uncertainties on each fraction are from the Binomial confidence interval. For sessions with one or two questions, the fraction of questions asked by women is low (~24%). For sessions with three to five questions, the fraction increases, and in sessions with six to nine questions, the fraction of questions from women is consistent with that the fraction of female speakers and attendees at the conference.

One solution to improve the inclusivity of question periods is to simply allow them to last longer. AAS talks are typically scheduled in 10-minute time-slots, including both the talk and the question period. This often leaves only 2–3 minutes for questions after each presentation. Such a short time is viewed as essential to allowing a large and diverse number of speakers to give talks during the meeting, but the consequence may be that men inevitably dominate the brief question sessions. If timing cannot be modified, persistent questioners could be encouraged to "share the air" (one of the many guidelines adopted during the 2015 Inclusive Astronomy Conference, available on their website at https://vanderbilt.irisregistration.com/Home/Site?code=InclusiveAstronomy2015) and allow others a chance to talk.

In a remarkable demonstration of gender conformity, the first author of this comment typically declines her opportunities to ask questions after conference talks, while the second author frequently raises his hand to participate. We both agree that other styles of scholarly interaction (e.g, Twitter exchanges, post-talk conversions, email, etc.) are valuable, but the public nature of question and answer periods elevates their importance. In large groups, frequency and duration of speech is an indicator of the perceived amount of power (Mast 2002). Women with some seniority often silence themselves, however, fearing backlash for speaking too much or having too much power – a fear that often proves to be true (Brescoll et al. 2011). When the first author chooses to ask fewer questions, is it a reflection of her personality, or a defense mechanism to fit into a culture that systematically devalues women?

**Towards a More Inclusive Question Survey**

The largest drawback of the gender question dataset is the binary treatment of gender, with men and women as two distinct groups and where gender is the proximate factor in asking a question. This reinforces a false gender binary, and excludes or inaccurately represents scientists who are non-binary and/or gender fluid, gender queer, or agender. The main hurdle in properly including all genders is in self-identification; the survey relies on volunteers gendering people from an observer's standpoint, and may be particularly prone to misgendering.

Gender is also not the sole determining factor in who asks questions at conferences. Seniority is frequently noted as a determining factor, which would skew the overall demographics as observed in our data. Other likely important factors that are not considered in our data are race, ethnicity, culture, national origin, gender self-identity, sexual orientation, disability, and the manner people who hold one or many of these identities interact with scientific culture; issues related to the intersection of identities in science are discussed in more detail in the comment by Dr. Chanda Prescot-Weinstein. A more comprehensive survey that includes self-identification of conference attendees should be adopted, but as the survey increases in complexity the involvement of trained sociologists or ethnographers, and dedicated data gathering personnel are desperately needed. We aim to not only know who is asking questions at astronomy meetings, but also how to include everyone in the conversation.

**Looking to the Future**

Previous Hack Day studies of the AAS survey data suggested that the gender of the first question asker could strongly influence the subsequent questions (Davenport et al. 2014). This indicated early participation of women in the Q/A sessions could encourage other women to ask subsequent questions. However, our analysis of the entire four-year AAS dataset found that while this trend is present, the effect is weak and may only have impact in longer Q/A sessions.

These types of subtle correlations in our data indicate that an ongoing survey has great potential to change how we conduct conference talks and question sessions. As a growing team of Hack Day participants, both young students and seasoned researchers alike, help analyze the data from the survey, we hope to ask new and more challenging questions of this data. This may reveal particular sessions or sub-fields that out-perform others in gender parity during question sessions; ultimately, this may be of use to conference organizers working to design inclusive question sessions.

Informally, we have heard feedback from some women that their knowledge of this project has encouraged them to actively participate in question sessions. While this effect biases our sample (towards a higher fraction of questions from women), we consider it a positive outcome, as our ultimate goal is to work towards more equitable and inclusive conference interactions. As such we will continue to call on the community to help gather gender survey data from future AAS meetings, and to initiate the survey at smaller subject-focused and regional meetings. Ideally, conference organizers and professional societies will adopt ours or similar methods, perhaps incorporating self-identification of identities beyond M/F gender, to both improve conferences in astronomy and work to become an example of inclusion at conferences across many fields.

Sarah J. Schmidt is at the Leibniz-Institute for Astrophysics Potsdam (AIP), An der Sternwarte 16, 14482, Potsdam, Germany. James R. A. Davenport at the Department of Physics & Astronomy, Western Washington University, Bellingham, WA 98225, USA. This work is supported by NSF Astronomy and Astrophysics Postdoctoral Fellowship under award AST-1501418

Email: sjschmidt@aip.de, james.davenport@wwu.edu